\documentstyle[aps,prl,twocolumn,epsfig]{revtex}
   
\begin{document}

\title{Penning collisions of laser-cooled metastable helium atoms}

\author{
F. Pereira Dos Santos, F. Perales$^{a}$, J. L\'eonard, A. Sinatra, Junmin Wang$^{b}$,\\
F. S. Pavone$^{c}$, E. Rasel$^{d}$, C.S. Unnikrishnan$^{e}$, M. Leduc\\
Laboratoire Kastler Brossel$^*$, D\'epartement de Physique, Ecole Normale Sup\'erieure, \\
24 rue Lhomond, 75231 Paris Cedex 05, France}                    

\date{}

\maketitle

\abstract{
We present experimental results on the two-body loss rates in a magneto-optical trap of
metastable helium atoms. Absolute rates are measured in a systematic way for several laser 
detunings ranging from -5 to -30 MHz and at different intensities, by monitoring the decay of
the trap fluorescence. The dependence of the two-body loss rate coefficient $\beta$ on the 
excited state ($2^3P_2$) and metastable state ($2^3S_1$) populations is also investigated. From these results we infer a rather uniform rate constant $K_{sp}=(1{\pm}0.4)\times10^{-7}$ cm$^3$/s.
\

PACS{
      {32.80.Pj}{Optical cooling of atoms;trapping}   \and
      {34.50.Rk}{Laser modified scattering and reactions}
     } 
}

\maketitle

\section{Introduction}
\label{intro}
Helium atoms in the metastable triplet state $2^3S_1$ (He*) appear to be a good candidate for Bose-Einstein Condensation (BEC) according to theoretical predictions \cite{Shlyap96}. 
The cross section for elastic collisions between spin-polarised metastable helium atoms is expected to be large, allowing efficient thermalization and evaporation in a magnetostatic trap, which is the standard technique to reach BEC \cite{Anderson95,Bradley957,Davis95,Fried98}.
On the other hand, very high autoionization rates (Penning collisions) prevent reaching high 
densities of metastable helium atoms, both in the presence and in the absence of light,  
unless the sample is spin polarized. \\
If a metastable helium atom collide either with an other metastable atom, or with an helium atom excited in the $2^3P_2$ state, the quasi molecule formed can autoionize according to 
the following reactions:

\begin{eqnarray}
{\rm He} \,(2^3S_1) + {\rm He} \,(2^3S_1) & \rightarrow & 
\left\{
\begin{tabular}{l}
	He\,$(1^1S_0)$ + He${}^{+}$ + $e^{-}$ \\
	He${}_2^+$  + $e^{-}$ 
\end{tabular} 
\right. 
\label{eq:ss}
\end{eqnarray}

\begin{eqnarray}
{\rm He} \,(2^3P_2) + {\rm He} \,(2^3S_1) & \rightarrow & 
\left\{
\begin{tabular}{l}
	He\,$(1^1S_0)$ + He${}^{+}$ + $e^{-}$ \\
	He${}_2^+$  + $e^{-}$ 
\end{tabular} 
\right.
\label{eq:sp} 
\end{eqnarray}

A first experiment at subthermal energy ($E=1.6$ meV) with the metastable helium system
was performed by M$\ddot{\rm{u}}$ller {\em et al.} \cite{Muller87}, allowing the determination of the interaction potentials.
Using those potentials 
the rate $\beta_{SS}$ for the reactions (\ref{eq:ss}) has been calculated 
\cite{Mast98,Kuma99,Vent99} to be a few $10^{-10}$ cm$^{3}$/s, which agrees with measurements 
performed in Magneto-Optical Traps (MOT) \cite{Bardou92,Mast98,Tol99}.
According to theoretical predictions \cite{Shlyap96}, the ionization rate corresponding to the reactions (\ref{eq:ss}) should be suppressed by four orders of magnitude in a magnetostatic trap.  Spin polarization of the atoms and spin
conservation in the collisional process are the causes of this suppression, which makes the quest of BEC reasonable. Actually, a reduction of more than a factor of 20 in the two-body loss rate in an optically polarized sample was observed experimentally \cite{Tol00}.  

In presence of light exciting the transition $2^3S_1 \rightarrow 2^3P_2$, the reaction
(\ref{eq:sp}) is dominant. ``Optical collisions" with metastable helium atoms were measured to 
have surprisingly large cross sections when compared with alkali systems \cite{Weiner99}. 
The study of optical collisions is of fundamental importance in order
to optimize the first step towards BEC, consisting in pre-cooling and 
trapping the atoms in a MOT. The goal is to transfer a cloud as dense as possible in a magnetic trap, in order to increase the elastic collision rate and start 
evaporation. The experimental study of optical collisions is the subject of this paper.

Several groups reported measurements of optical collisions rates, by studying losses in the 
MOT at small detunings \cite{Bardou92,Kuma99,Tol99} around $-5$ MHz and at large
detunings \cite{Tol99} at $-35$ MHz and $-45$ MHz. 
Measurements over a broad range of detunings, from $-5$ MHz to $-20$ MHz, were reported in 
\cite{Brow00} and the dependence of the loss rate on the intensity of the MOT laser beams was 
investigated.
In reference \cite{Mast98} a theoretical model for optical collisions is also proposed predicting 
rates in good agreement with the measurements, but differing by more than one order of 
magnitude with all the other measurements previously quoted. 

Our measurements are performed in a MOT loaded with $10^9$ atoms, at a peak density of $10^{10}$ atoms/cm$^{3}$. 
With respect to previous works, we extend the measurements of the two-body loss rate 
to a wider range of detunings and intensities with a good precision, by measuring the number of atoms and the size of the trap using absorption techniques. 
Also, by measuring accurately the excited state population in each trapping condition, we are able
to interpret our data with a simple model, expressing the two-body loss rate
in terms of the excited state population and of a rate constant $K_{sp}$, found to be 
independent of the laser detuning and intensity.

Our experimental setup is described in section \ref{sec:setup}, while in
section \ref{sec:detection} we explain our detection system and we give the
working conditions and performance of our magneto-optical trap. 
In section \ref{sec:fluo} we describe in detail the experimental procedure used to measure the two-body loss rate and the excited state population for different
trapping conditions. The results are given in section \ref{sec:results}, and the conclusions
in section \ref{sec:concl}.

\section{Experimental set-up}
\label{sec:setup}

A beam of metastable helium atoms is generated by a continuous high voltage discharge in helium gas, cooled to liquid nitrogen temperature. 
Radiation pressure on the metastable beam allows one to increase its brightness, and to deflect it from the ground state helium beam \cite{Rasel99}. 
The metastable atoms are then decelerated by the Zeeman slowing technique 
and loaded in a magneto-optical trap (MOT) in a quartz cell at a background pressure of $5\times10^{-10}$ torr.  
More details on the experimental setup will be given in a forthcoming paper \cite{Pereira00}.
MOT parameters for optimal loading of the trap are listed in table \ref{tab:parMOT}.
For the laser manipulation of the atoms, we use the line at 1083 nm, 
connecting the metastable triplet state $2^{3}S_1$ to the radiative state $2^{3}P_2$. 
The saturation intensity $I_{sat}$ for this transition is 0,16 mW/cm$^{2}$ and the 
linewidth $\Gamma/2\pi$ is 1.6 MHz. 
Our laser system consists of a DBR laser diode (SDL-6702-H1) in an extended cavity configuration, 
injecting a commercial Ytterbium doped fiber amplifier (IRE-POLUS Group).
The diode is stabilized by saturation spectroscopy at -240 MHz from resonance. At the fiber output we obtain 600 mW of power, in a TEM00 mode at the same frequency.  
The estimated linewidth is around 300 kHz. All the frequencies required for
collimation, deflection, trapping and probing are generated by acousto-optical modulators in a 
double pass configuration, while we use directly part of the fiber output beam for slowing
the atoms.
\begin{table}[htb]
\caption{Optimal loading parameters of the He$^{*}$ magneto-optical trap.}
\label{tab:parMOT}
\vspace{0.1in}
\begin{center}
\begin{tabular}{|l|l|}
\hline
Laser detuning &   -45MHz  \\ \hline 
Laser beam diameter&            2 cm\\ \hline
Vertical laser intensity (Ox)&       2$\times$9 mW/cm$^{2}$\\ \hline
Longitudinal laser intensity (Oy)&            2$\times$9 mW/cm$^{2}$\\ \hline
Transverse laser intensity (Oz)&           2$\times$7 mW/cm$^{2}$\\ \hline
Total intensity&                        50 mW/cm$^{2}$\\       \hline 
Weak axis magnetic field gradient&               $b_x=b_y=20$ G/cm \\  \hline
Strong axis magnetic field gradient&              $b_z=40$ G/cm\\  \hline
\end{tabular}
\end{center}
\end{table}
\section{Detection system and characterization of the MOT}
\label{sec:detection}
\begin{figure}[htb]
\begin{center}
\epsfig{file=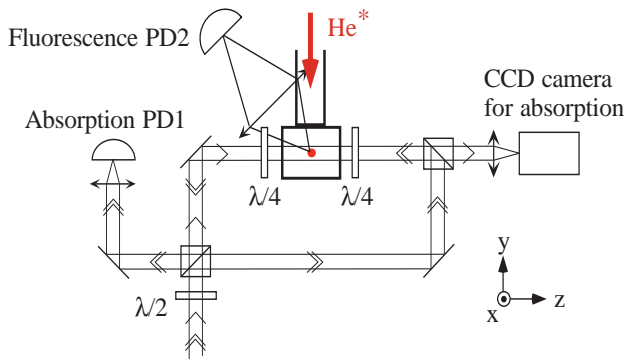,height=4.7cm,width=8.5cm}
\end{center}
\caption{\footnotesize Detection set-up. By rotating the $\lambda$/2 plate, one can create either a progressive plane wave for measuring the absorption on the photodiode PD1, or a standing wave, with both beams circularly polarized in the cell region, for imaging the cloud onto the CCD camera. PD2 monitors the fluorescence of the MOT.}
\label{detec}
\end{figure}
\begin{figure}[htb]
\begin{center}
\epsfig{file=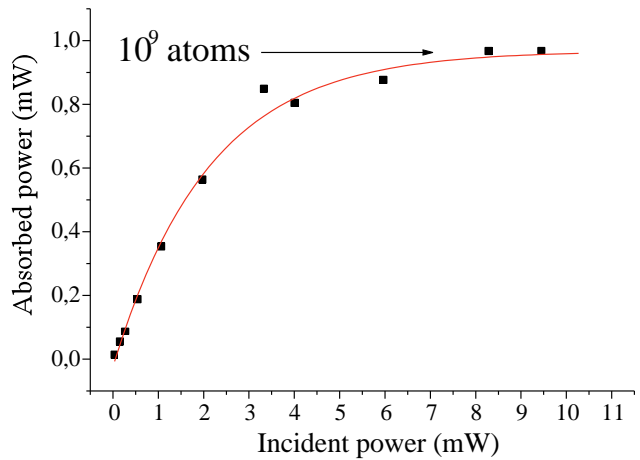,height=6cm,width=8.5cm}
\end{center}
\caption{\footnotesize Absorbed power by the MOT versus incident power of the laser probe beam. The absorbed power saturates at 1 mW for an incident power of 10 mW. The corresponding number of atoms is $(1{\pm}0.1)\times10^{9}$ atoms}
\label{abs}
\end{figure}

In order to fully characterize the cloud, we use a probe laser beam on resonance, whose diameter is about 1 cm, which is turned on 100 $\mu$s after the MOT field and light beams have been turned off.
Our detection setup (see fig. \ref{detec}) allows different measurements. 
With the combination of $\lambda$/2 plates and polarization beam splitter cubes, we can create 
either (i) a progressive wave, circularly polarized, passing through the atomic 
cloud towards a photodiode (PD1 in figure \ref{detec}), giving the total absorption by the 
atoms, or (ii)  a stationary wave, 
also circularly polarized, one arm of which is sent to a 
CCD camera, allowing spatially resolved absorption pictures of the cloud. 
A second photodiode (PD2 in figure \ref{detec}) is used to collect the cloud fluorescence. 
We use the absorption photodiode PD1 to measure N, the number of atoms in the steady state of the MOT. 
The probe beam saturates the transition when the incident power
exceeds 10 mW (see fig. \ref{abs}). The maximum absorbed power is then 
$P=N h \nu \frac{\Gamma}{2}$. Our Watt-meter (Coherent lab-master) is calibrated to 3\% accuracy and allows a rectilinear calibration fit of the photodiode voltage. 
We measure a maximum total absorption of 1 mW, corresponding to $(1{\pm}0.1)\times10^{9}$ atoms. 
We estimate the accuracy for the measurement of N to be about 10\%.\\
The typical parameters of our magneto-optical trap with the operating conditions of table \ref{tab:parMOT} are listed in table \ref{tab:resMOT}.

\begin{table}[htb]
\caption{Characterization of the MOT with parameters of table \ref{tab:parMOT}.}
\label{tab:resMOT}
\vspace{0.1in}
\begin{center}
\begin{tabular}{|l|l|}
\hline
Number of atoms &    $N=(1{\pm}0.1)\times10^9$   \\ \hline
RMS size (weak axis) &  $\sigma_x=\sigma_y=(2{\pm}0.1)$ mm  \\ \hline
RMS size (strong axis) &  $\sigma_z=(1.6{\pm}0.1)$ mm  \\ \hline
Density at the center &       $(1{\pm}0.25)\times10^{10}$ atoms/cm$^3$ \\ \hline
Temperature &              $1$ mK\\  \hline
\end{tabular}
\end{center}
\end{table}

We stress the fact that the case of He* differs of that of alkalis, for which the imaging method gives a direct measurement of both the two-dimensional column density and the rms sizes of the MOT, by absorption of a brief and low intensity probe pulse ($I\ll I_{sat}$). In the case of He*, the quantum efficiency of the CCD camera ($10^{-3}$ at 1.083 $\mu m$) is too low to provide images with a sufficient signal to noise ratio. We need instead to illuminate the atoms with a 200 $\mu s$ pulse whose intensity is about 
0.1 mW/cm$^{2}$ ($I\sim I_{sat}$), and use a moderate magnification of 1/5.  
Another difficulty with He* occurs from the large recoil momentum $\hbar k/m$ (9.2 cm/s) due to the light mass of the atoms : the atoms are pushed out of 
resonance during the 200 $\mu s$ pulse if a traveling wave pulse is used. 
The solution we adopted is to illuminate the atoms in a standing wave with the set-up shown in figure
\ref{detec}. Though this scheme allows us to obtain pictures with a good 
contrast, 
the drawback is that the images obtained in the standing wave 
configuration for $I\sim I_{sat}$ are more difficult to analyze than in the low intensity case.
In order to interpret the absorption pictures in the standing wave configuration, and for any
saturation parameter, we developed a handy theoretical model (see appendix \ref{app:model})
giving the column density of the atoms for each pixel of the CCD camera. The resulting density 
is then fitted by a Gaussian curve to extract the size of the cloud.\\

\section{Measuring the trap decay by fluorescence}
\label{sec:fluo}
\begin{figure}[htb]
\begin{center}
\epsfig{file=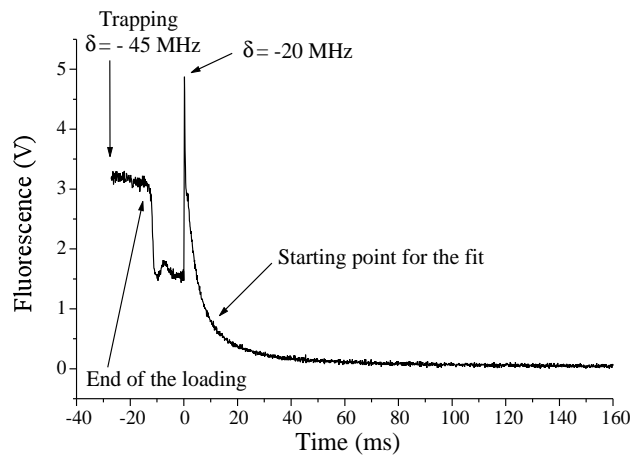,height=6cm,width=8.5cm}
\end{center}
\caption{\footnotesize Evolution of the fluorescence signal. Once the loading is stopped, scattered light from the slowing beam is blocked, which explains the drop of the signal at t=-10 ms. The detuning is then set to $\delta = -20$ MHz at t=0 ms and the fluorescence decays.}
\label{decay}
\end{figure}

\begin{figure}[bht]
\begin{center}
\epsfig{file=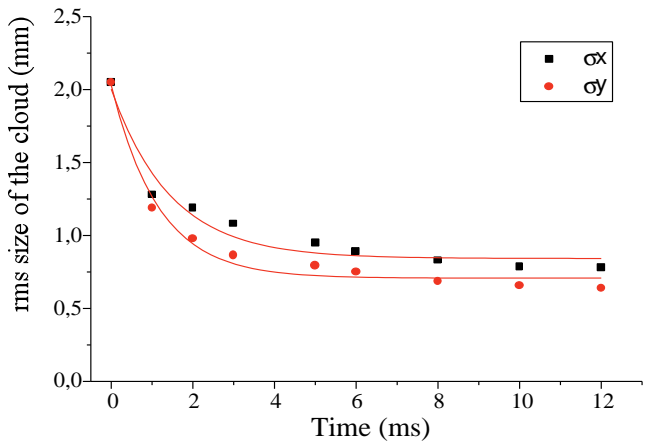,height=6cm,width=8.5cm}
\end{center}
\caption{\footnotesize Size of the MOT during the compression phase. The new equilibrium is reached after 10 ms}
\label{comp}
\end{figure}

\begin{figure}[htb]
\begin{center}
\epsfig{file=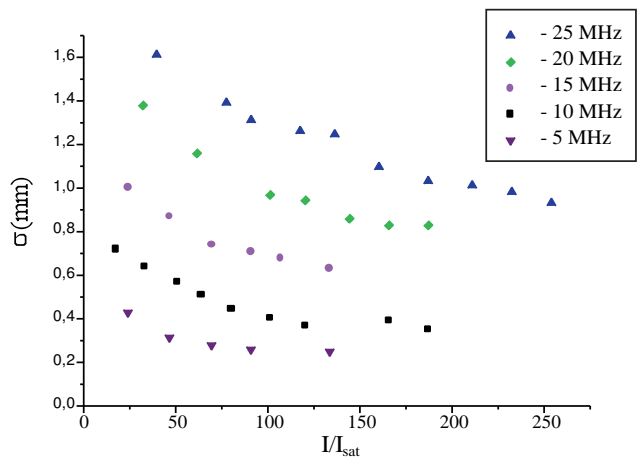,height=6cm,width=8.5cm}
\end{center}
\caption{\footnotesize Rms size of the MOT cloud as a function of the intensity of the MOT laser beams for various detunings.}
\label{size}
\end{figure}

\begin{figure}[htb]
\begin{center}
\epsfig{file=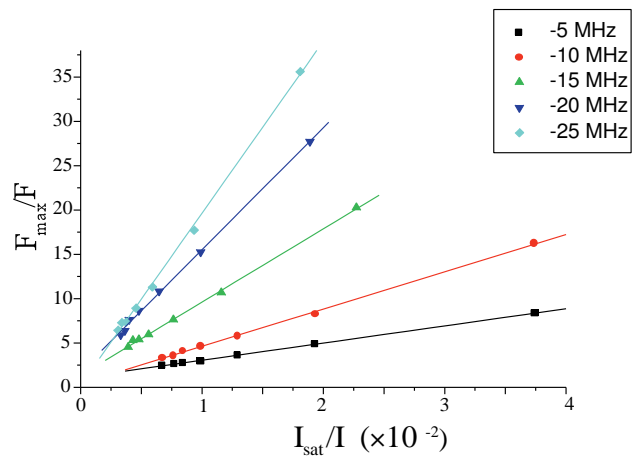,height=6cm,width=8.5cm}
\end{center}
\caption{\footnotesize Fluorescence signal F from the MOT as a function of intensity I of the laser beams. The inverse of the fluorescence $F$ is found to vary linearly with the inverse of the intensity $I$. The results are used for the calibration of the number of atoms.}
\label{calfluo}
\end{figure}

Once the loading of the MOT is interrupted, the evolution of the number 
of trapped atoms $N$ is given by the following equation: 
\begin{equation}
\frac{dN}{dt}=-\alpha N-\beta \int n^{2}({\bf r},t)d^{3}{\bf r}  
\end{equation}
where $n({\bf r},t)$ is the atomic density at position {\bf r} and time t, $\alpha$ is the decay rate due to collisions between trapped atoms and the residual gas, 
and $\beta$ is the two body intra-MOT loss factor.  
Assuming that the spatial distribution is independent of the time evolution of 
the number of atoms, which is valid at low enough densities, one can write the density as 
\begin{equation}
n({\bf r},t)=\frac{N(t)}{(2 \pi)^{\frac{3}{2}} 
\sigma_x \sigma_y \sigma_z}e^{- \frac{x^{2}}{2 \sigma_x^{2}} - 
\frac{y^{2}}{2 \sigma_y^{2}} - \frac{z^{2}}{2 \sigma_z^{2}}}  
\end{equation}
At low enough pressure and high enough density, losses due to background gas are negligible, 
so that the equation reduces to
\begin{equation}
\frac{dN}{dt}=-\beta \frac{N^{2}(t)}{(4 \pi)^{\frac{3}{2}} \sigma_x \sigma_y \sigma_z}  
\end{equation}
whose solution is 
\begin{equation}
N(t)=\frac{N(t_0)}{1+\frac{\beta}{2 \sqrt{2}} n({\bf 0},t_0) (t-t_0)} \label{eq:Nt} 
\end{equation}
where $t_0$ is the initial time.
In order to follow the evolution of the number of trapped atoms, we monitor the fluorescence decay 
of the MOT with a photodiode (PD2 in figure \ref{detec}). 
As the fluorescence signal is proportional to the number of atoms, 
we obtain a fluorescence decay curve reproducing equation (\ref{eq:Nt}), 
which we fit to get the parameter $\beta n({\bf 0},t_0)$. 
In order to determine $\beta$, one still has to measure $n({\bf 0},t_0)$, 
which means that one has to measure the rms size of the cloud along the three directions and the initial
number of atoms $N(t_0)$.\\
Our goal is to measure the loss rate for a wide range of detunings and intensities. The experimental procedure, divided in three successive steps, is the following.\\
(1) First, we load the trap for 1 s at $\delta = - 45$ MHz and at the highest intensity in the trapping beams (I/Isat=50 per laser arm). Then, we stop the loading by blocking the slowing beam with a mechanical shutter. 20 ms later, we ``compress" the MOT by suddenly changing its detuning and 
intensity using acousto-optical modulators. We record the fluorescence signal during this procedure. A typical fluorescence curve is shown in figure \ref{decay}.
The loading is stopped at t=-20 ms and the photodiode signal drops by a factor of 2 at t=-10 ms because the background light from the slowing beam is blocked. The fluorescence is greatly enhanced in the
beginning of the compression phase at t=0 ms, 
as expected when the detuning is set closer to resonance (the detuning is set here to -20 MHz), 
but decays to almost zero in about 100 ms because of the two-body losses.
Figure \ref{comp} shows the time evolution of the size of the cloud during this phase of compression, showing that 10 ms are enough to reach the new equilibrium size. 
Thus, we extract the parameter $\beta n({\bf 0},t_0)$ from a fit of the fluorescence decay starting 
from $t=t_0=10$ ms. At this very time we measure the sizes of the MOT along x and y and the number
of atoms in order to calculate  $n({\bf 0},t_0)$.\\ 
(2) Then, the sizes along the weak axis of the magnetic field gradient are measured by absorption on the CCD camera as explained in section \ref{sec:detection}.
Figure \ref{size} shows the rms size along x for various laser detunings and intensities. 
The size along z (strong axis of the quadrupole field) is inferred from measurements of the 
sizes along x and y with a magnetic field gradient $b$ twice as large. 
We find  a typical size along z 20\% smaller than along the weak axes of the quadrupole.  
We did not correct the sizes for the expansion of the cloud during the pulse lasting 
200 $\mu s$, as this would have required the measurement of the temperature for all the detunings and intensities. Nevertheless, we performed some time of flight measurements, giving temperatures ranging from 0.3 mK at -10 MHz to 1 mK at -40 MHz, from which we estimate that the sizes are overestimated at most by 5 \% at -25 MHz and by 15 \% at -5 MHz. 
In addition, we measured the statistical error on the sizes to be relatively small at large detunings, 2 to 3\%, but larger at small detunings (about 10\% at -5 MHz). This is due (i) to the poor spatial resolution of our imaging system (pixel dimension $80 \mu m\times130 \mu m$), and (ii) to a low signal to noise ratio for small detunings where the loss rate is larger, as most of the atoms are lost during the compression phase.
(3) Finally, to determine the number of atoms that were still trapped at $t_0=10$ ms, we simultaneously switch off the magnetic field and set the trapping beams on 
resonance at $t_0$, instead of letting the trap decay as in figure \ref{decay}. The laser intensity is set to a high enough value to strongly saturate the transition. 
We get a peak of fluorescence, whose amplitude is proportional to the number of atoms. We compare it with the peak obtained with the same procedure but for the MOT in the best loading conditions of figure \ref{abs}, for which we measured the number of atoms precisely. 
From this comparison, we infer the number of atoms at $t=t_0$ in the compressed MOT, and thus determine $n({\bf 0},t_0)$.\\
This measurement also gives access to the value of the average population of the excited state $\pi_p$. Indeed, $\pi_p$ is given by
\begin{equation}
\frac{F}{F_{max}}=\frac{\pi_p}{1/2}=2\times\pi_p
\label{eq:pip} 
\end{equation}
where $F$ is the fluorescence signal we measure in the compressed MOT at $t_0$, and $F_{max}$ the fluorescence signal at resonance, when the transition is saturated, and $\pi_p$ expected to be 1/2.\\
Figure \ref{calfluo} shows the results of the fluorescence measurements, giving $F_{max}/F$ as a function of the inverse of the laser intensity I for various detunings.
It is interesting to note that the inverse of $F$ is found to vary linearly with the inverse of $I$.
Following \cite{Town95}, the fluorescence of N atoms in the compressed MOT can be modeled by the following equation:
\begin{equation}
F=\eta \, N \, h \nu \, \frac{\Gamma}{2}\, 
\frac{C_1\frac{I}{I_{sat}}}{1+C_2\frac{I}{I_{sat}}+4\frac{\delta^{2}}{\Gamma^{2}}}\ 
\label{fluo}
\end{equation}
where $\eta$ is the detection efficiency, $I$ is the total intensity of the six MOT beams,
and $C_1$ and $C_2$ phenomenological factors. $C_1$ and $C_2$ would be 1 for a two-level atom, but they are expected to be smaller for an atom placed at the intersection of 6 differently polarized laser beams, as happens in a MOT. In reference \cite{Town95}, $C_1$ and $C_2$ are found to be equal, and slighly larger than the average of the squares of the Clebsch-Gordan coefficients over all possible transitions. For a $J=1 \longleftrightarrow J=2$ transition, this average is 0.56.  
We can rewrite equation (\ref{fluo}) as
\begin{equation}
\frac{F_{max}}{F} =\frac{C_2}{C_1}+\frac{1+4\frac{\delta^{2}}{\Gamma^{2}}}{C_1} \frac{I_{sat}}{I}\ 
\label{fluo2}
\end{equation}
where $F_{max}=\eta N h \nu\frac{\Gamma}{2}$.\\
The results of figure \ref{calfluo} show a good agreement with (\ref{fluo2}). But, $C_2$ and $C_1$ are not found equal, and both depend on the detuning. For example, $C_1$ is found to be 0.58, 0.48, 0.46, 0.44, 0.22 for $\delta=$-25, -20, -15, -10, -5 MHz respectively. We stress the fact that, for the fluorescence at resonance, and for full saturation, C1 and C2 are expected to be equal.

\section{Results}
\label{sec:results}
\begin{figure}[htb]
\begin{center}
\epsfig{file=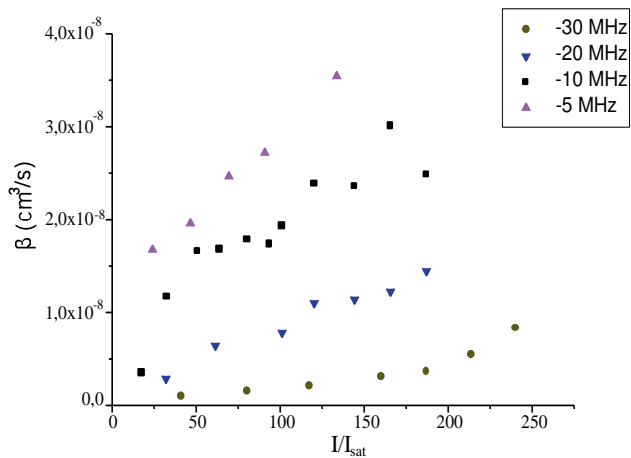,height=6cm,width=8.5cm}
\end{center}
\caption{\footnotesize Two-body loss rate factor as a function of laser power for several detunings.}
\label{beta}
\end{figure}

\begin{figure}[htb]
\begin{center}
\epsfig{file=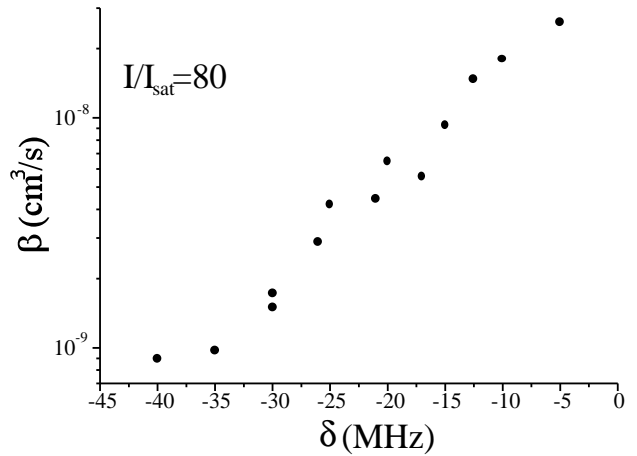,height=6cm,width=8.5cm}
\end{center}
\caption{\footnotesize Two-body loss rate factor as a function of detuning for a fixed intensity ${I}=80{I_{sat}}$ of the laser.}
\label{betadelta}
\end{figure}

The results of the Penning collisions rate ${\beta}$ are shown in
figures \ref{beta} and \ref{betadelta}.\\
Figure \ref{beta} presents the loss parameter ${\beta}$ as a function of the laser intensity for different detunings $\delta$, from -30 to -5 MHz. The uncertainty of the measurements varies from 25 $\%$ for large detunings to 60 $\%$ for small detunings. For all detunings, $\beta$ increases with power, which shows that S-P collisions are dominant.\\  
Fig. \ref{betadelta} shows the loss parameter as a function of detuning for a fixed intensity ($\frac{I}{I_{sat}}=80$). For the same reason, the rate increases when the detuning goes to zero, as the population in the P state increases. 
Our results for $\beta$ agree with previous results 
\cite{Bardou92,Kuma99,Tol99,Brow00} within the given error bars, extending the measurements to a wider range of parameters. 
For example, at -5 MHz and in an intensity range for which $\beta$ is not expected to
vary strongly ($I=140$ to $200 I_{sat}$), Kumakura et al. \cite{Kuma99} find $\beta=(4.2{\pm}1.2)\times10^{-8}$ cm$^{3}$/s, 
Browaeys et al. \cite{Brow00} $\beta=2\times10^{-8}$ cm$^{3}$/s with an uncertainty of a factor 2 and Tol et al. \cite{Tol99} $\beta=(1.3{\pm}0.3)\times10^{-8}$ cm$^{3}$/s. Our measurement $\beta=(3.5{\pm}1.4)\times10^{-8}$ cm$^{3}$/s agrees best with \cite{Kuma99}. 
One should also note that we find neither a decrease of $\beta$ for high intensities at small 
detunings, nor a decrease of $\beta$ at small detunings for a given intensity : this differs from the results of \cite{Brow00}. In fact, we find that $\beta$ increases with intensity at small detunings, and also increases with decreasing detunings at a given intensity. We also disagree with the results of \cite{Mast98} where much smaller rates are found.\\
Finally, we also measured the loss rate in the trapping conditions ($\delta=-45$ MHz, $I=310I_{sat}$) : the decay rate  of the number of atoms was found to be $\beta n({\bf 0})=30$s$^{-1}$ at a density of $10^{10}$ at/cm$^3$, which gives $\beta=3\times10^{-9}$ cm$^3$/s.\\
One can further analyze these data following the simple model of \cite{Bardou92} which relates the decay constant $\beta$ to the constant rate coefficients $K_{ss}$, $K_{sp}$ and $K_{pp}$ and to the populations of the excited and ground state levels, $\pi_p$ and $\pi_s$ respectively:
\begin{equation}
\beta = K_{ss} \, \pi_s \, \pi_s + 2 K_{sp} \, \pi_s \, \pi_p + K_{pp} \, \pi_p \, \pi_p
\end{equation}
Experiments \cite{Mast98,Tol99} or theory \cite{Mast98,Kuma99,Vent99} have shown that the contributions $K_{ss} \pi_s^2$ and $K_{pp} \pi_p^2$ to the total rate $\beta$ 
are smaller than the $K_{sp}$ term by approximately two orders of magnitude.\\
From the measurements of the fluorescence signal in figure \ref{calfluo}, we derive $\pi_p$
for each experimental point, as $F/F_{max}$ in eq. (\ref{eq:pip}) is equal to $2\times\pi_p$.
\begin{figure}[htb]
\begin{center}
\epsfig{file=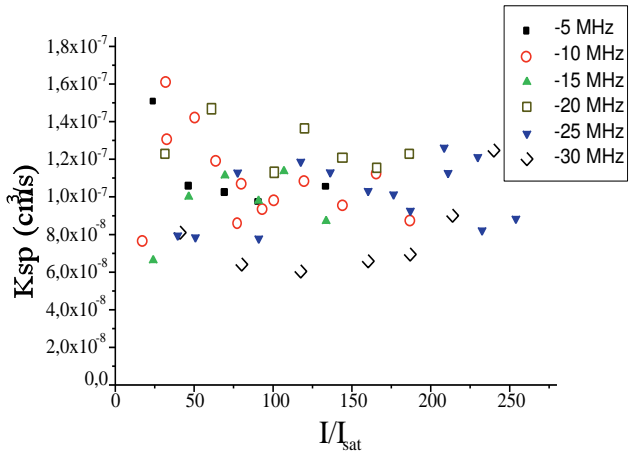,height=6cm,width=8.5cm}
\end{center}
\caption{\footnotesize Rate coefficient $K_{sp}$ for all our measurements, as a function of the laser intensity I for several detunings.}
\label{fig:ksp}
\end{figure}
In figure \ref{fig:ksp}, we then plot $K_{sp}$ for the ensemble of our data.
We do not see clear evidence for a dependence of $K_{sp}$ with the detuning or the intensity within
the dispersion of our data. To a good approximation, we estimate then that $K_{sp}$
is actually constant in the explored range of parameters: $K_{sp}=(1.0{\pm}0.4)\times10^{-7}$ $\mbox{cm}^3$/s, with a dispersion that roughly agrees with the error bars we claim. This result agrees with the first measurement ever performed \cite{Bardou92}, but the precision is now much improved.
It also agrees well with the measurements of \cite{Kuma99} where the authors found $K_{sp}=(8.3{\pm}2.5)\times10^{-8}$ $\mbox{cm}^3$/s, assuming that for their parameters ($\delta=-5$ MHz and $I=30$ mW/cm$^2$), $\pi_s=\pi_p=0.5$.\\
An important point is that, in contrast with the measurement of the fluorescence at resonance where the transition is assumed to be saturated, $\pi_p$ in the compressed MOT never reaches 0.5 in our measurements : even for the 
smallest detuning and the highest intensity, $\pi_p$ is only 0.2.
This explains why the results for $\beta$ in figure \ref{beta} at $\delta=-5$ MHz 
strongly increase for increasing intensity over the whole explored range. 

\section{Conclusion}
\label{sec:concl}
We measured the absolute two-body loss rate between metastable atoms in a magneto-optical trap as a function of detuning and intensity. We extended the range of these parameters and compared the results to those of previous measurements, mostly performed at small detunings. Using a new experimental approach, we obtained reliable values for the two-body loss rates with an improved accuracy as compared to most earlier results. In the region of overlap of parameters, we find a good agreement with previous measurements, within the quoted uncertainties. We find a loss rate monotonically increasing as a function of intensity and decreasing with detuning.
Our measurements are interpreted with a simple model, giving a rather constant loss rate $K_{sp}$, with an average value of $(1{\pm}0.4)\times10^{-7}$ cm$^3$/s, as already found in the very first measurement of \cite{Bardou92}. We believe that the quality and the extended range of our measurements should motivate more theoretical work, in order to understand better the peculiar dynamics of Penning collisions between metastable helium atoms in the presence of light.\\

\noindent {\bf Acknowledgments:}
The authors wish to thank C. Cohen-Tannoudji for helpful discussions and careful reading of the manuscript.\\

$^a$ Permanent address: Laboratoire de Physique des Lasers, UMR 7538 du CNRS, Universit\'e Paris Nord, Avenue J.B. Cl\'ement, 93430 Villetaneuse, France.

$^b$ Permanent address: Institute of Opto-Electronics, Shanxi University, 36 Wucheng Road, Taiyuan, Shanxi 030006, China.

$^c$ Permanent address : Dept. of Physics, Univ. of Perugia, Via Pascoli, Perugia, Italy; Lens and INFM, L.go E. Fermi 2, Firenze, Italy

$^d$ Present address : Universit$\ddot{\rm{a}}$t Hannover, Welfengarten 1, D-30167 Hannover, Germany.

$^e$ Permanent address : TIFR, Homi Bhabha Road, Mumbai 400005, India.

$^*$ Unit\'e de Recherche de l'Ecole Normale Sup\'erieure et de
l'Universit\'e Pierre et Marie Curie, associ\'ee au CNRS (UMR 8552).

\appendix
\section{Model of the absorption}
\label{app:model}
In this appendix we describe the method we used to quantitatively interpret the 
absorption images of the atomic cloud when a standing wave configuration of the probe beam
is used, and for an arbitrary saturation parameter.
We describe the atoms as two-level atoms characterized by a non linear susceptibility:
\begin{equation} 
\chi = n(x,y,z) \, \left[ - \frac{d^2}{\hbar \epsilon_0} \;\;
 		  \frac{\delta - i (\Gamma/2)}{(\Gamma/2)^2+\delta^2+|\Omega|^2/2} \right]
\label{eq:chi}
\end{equation} 
where $n(x,y,z)$ is the atomic density, 
$d$ the atomic dipole, $\delta$ the detuning, $\Gamma$ the inverse
lifetime of the excited state and $\Omega$ is the Rabi frequency given by
\begin{equation}
\frac{\hbar \Omega}{2}= -d \, {\cal{E}}^{(+)} \hspace{1cm} 
		E = {\cal{E}}^{(+)} \, e^{-i\omega t} + c.c. 
\end{equation}
where E is the electric field.
The direction of propagation of the beam is $z$ and the field is supposed to be uniform 
in the plane (x,y). 
The propagation of the field is then described by the Maxwell equations:
\begin{equation} 
\left[ \Delta + k_0^2 \, (1+\chi) \right] \Omega(z) = 0
\label{eq:prop}
\end{equation} 
where $k_0$ is the wavevector of the light. \\
The principle of the model is to use the slowly varying envelope approximation generalized to the case
of a standing wave. We then decompose the probe beam field as:
\begin{equation} 
\Omega(z) = A_+(z) \, e^{i k_0 z} +  A_-(z) \, e^{- i k_0 z}
\label{eq:omegaz}
\end{equation} 
where $A_+$, $A_-$ are the slowly varying envelopes of the wave going towards
positive $z$ and negative $z$ respectively. A similar decomposition holds for
the nonlinar susceptibility of the atoms:
\begin{equation} 
\chi(z) = \chi_0(z) + \chi_+(z) \, e^{2 i k_0 z} +  \chi_-(z) \, e^{- 2 i k_0 z} + \ldots
\label{eq:chiz}
\end{equation} 
where $\chi_0$, $\chi_+$ and $\chi_-$ are slowly varying envelopes, and where we neglect
terms in the expansion describing generation of frequencies others than the probe
frequency via the non linear interaction. \\
If we insert the expansions (\ref{eq:omegaz}) and  (\ref{eq:chiz}) into the propagation
equation (\ref{eq:prop}) and use the rotating wave appoximation, we obtain a set of
two coupled differential equations for the slowly varying field amplitudes $A_+$, $A_-$. 
By splitting the complex amplitudes into modulus and phase:
\begin{equation}
A_+ = |A_+| \, e^{i \phi_+}  \hspace{1cm}
A_- = |A_-| \, e^{i \phi_-}
\end{equation}
and by introducing the phase difference $(\phi_+ - \phi_-)$ in the definition of
the slowly varying susceptibilities $\chi_+$ and $\chi_-$:
\begin{equation}
\chi_{+} = \tilde{\chi}_{+} \, e^{i \, (\phi_+ - \phi_-)} \hspace{1cm}
\chi_{-} = \tilde{\chi}_{-} \, e^{- i \, (\phi_+ - \phi_-)}  \,,
\end{equation}
one can write : 
\begin{eqnarray}
\frac{d |A_+|}{dz} & = & \frac{k_0}{2} \left( {\rm Im} \tilde{\chi}_+ \, |A_-| +
					      {\rm Im} \tilde{\chi}_0 \, |A_+| \right)\\
\frac{d |A_-|}{dz} & = & - \frac{k_0}{2} \left( {\rm Im} \tilde{\chi}_- \, |A_+| +
					        {\rm Im} \tilde{\chi}_0 \, |A_-| \right) \,.
\end{eqnarray}
By using expressions (\ref{eq:chi}) and (\ref{eq:omegaz}), the quantities 
$k_0 \, {\rm Im} \tilde{\chi}_+$, $k_0 \, {\rm Im} \tilde{\chi}_-$
and $k_0 \, {\rm Im} \tilde{\chi}_0$ are readily calculated:
\begin{eqnarray}
k_0 \, {\rm Im} \tilde{\chi}_0 &=& \frac{3 \lambda^2}{2 \pi}\,  n(x,y,z) \, \alpha \, f_0 \\
k_0 \, {\rm Im} \tilde{\chi}_+ &=& k_0 \, {\rm Im} \tilde{\chi}_-
					= \frac{3 \lambda^2}{2 \pi} \, n(x,y,z) \, \alpha \, f_1 
\end{eqnarray}
where
\begin{eqnarray}
\alpha & = & \frac{(\Gamma/2)^2}{(\Gamma/2)^2+\delta^2+
	\frac{1}{2}\left( |A_+|^2+|A_-|^2 \right) }\\
f_0 & = & \frac{1}{\sqrt{1-\epsilon^2}} \;;\hspace{1cm} f_1  =  \frac{1-f_0}{\epsilon} \\
\epsilon & = & \frac{|A_+||A_-|}{(\Gamma/2)^2 + \delta^2 +
	\frac{1}{2}\left( |A_+|^2+|A_-|^2 \right) } \,.
\end{eqnarray}
As a last step we eliminate the atomic density $n(x,y,z)$ from the equations by changing
variable:
\begin{equation}
Z(z)=\int_{-\infty}^z n(x,y,z^\prime) dz^\prime
\end{equation}
and we obtain the final coupled equations:
\begin{eqnarray}
\frac{d |{\tilde{A}}_+|}{dZ} & = & \frac{3 \lambda^2}{4 \pi} \, \alpha \, 
	\left( f_1 |{\tilde{A}}_-| + f_0 |{\tilde{A}}_+| \right) \label{eq:final1}\\
\frac{d |{\tilde{A}}_-|}{dZ} & = & \frac{3 \lambda^2}{4 \pi} \, \alpha \, 
	\left( f_1 |{\tilde{A}}_+| + f_0 |{\tilde{A}}_-| \right) \label{eq:final2}\,,
\end{eqnarray}
where:
\begin{equation} 
\tilde{A}_-= A_-/(\Gamma/2) \hspace{1cm}  \tilde{A}_+= A_+/(\Gamma/2) \,.
\end{equation} 
For $\delta=0$ and in the limit of small saturation parameters, one has
$\alpha=1$, $f_0 \simeq 1$, $f_1 \simeq 0$ and one recovers
the usual decoupled equation for low saturation absorption.
We have now to solve equations (\ref{eq:final1}) and (\ref{eq:final2}).
More precisely we wish to calculate the column density 
\begin{equation}
Z^{\infty} = \int_{-\infty}^{+\infty} n(x,y,z^\prime) dz^\prime
\label{eq:column}
\end{equation}
for each effective pixel (x,y) of our image of the cloud. 
For each effective pixel, we can measure the initial conditions:
\begin{eqnarray}
|{\tilde{A}}_+|^2(Z(-\infty)=0) & = & I_{i} \label{eq:init1} \\
|{\tilde{A}}_-|^2(Z(-\infty)=0) & = & I_{f} \label{eq:init2}
\end{eqnarray}
corresponding respectively to the intensity of the probe beam before passing through the cloud, or equivalently without the atoms, and  
to the intensity of the probe beam that passed through the atomic cloud.
For symmetry reasons, the column density (\ref{eq:column}) is given by $2 Z_0=Z(0)$, where $Z_0$ verifies
\begin{equation}
|{\tilde{A}}_+(Z_0)|^2=|{\tilde{A}}_-(Z_0)|^2.
\end{equation}
For each pixel (x,y), we then
integrate equations (\ref{eq:final1})-(\ref{eq:final2}) numerically
using the initial conditions (\ref{eq:init1})-(\ref{eq:init2}) until $|{\tilde{A}}_+(Z)|^2=|{\tilde{A}}_-(Z)|^2$. The corresponding $Z$ multiplied
by 2 gives the column density.
Note that, contrarily to what happens in the low saturation regime,
we here need the values $I_{i}$ and $I_{f}$ separately (and not only their ratio),
which implies a calibration of our CCD camera.

\end{document}